# High-throughput direct measurement of magnetocaloric effect based on lock-in thermography technique

Yusuke Hirayama,[a)] Ryo Iguchi, Xue-Fei Miao,[b)] Kazuhiro Hono, and Ken-ichi Uchida[c)]

*National Institute for Materials Science, Tsukuba 305-0047, Japan.*

**ABSTRACT**

We demonstrate a high-throughput direct measurement method for the magnetocaloric effect (MCE) by means of a lock-in thermography (LIT) technique. This method enables systematic measurements of the magnetic-field and operation-frequency dependences of the temperature change induced by the MCE. This is accomplished in a shorter time compared with conventional adiabatic temperature measurement methods. The direct measurement based on LIT is free from any possible miscalculations and errors arising from indirect measurements using thermodynamic relations. Importantly, the LIT technique makes simultaneous MCE measurements of multiple materials possible without increasing measurement time, realizing high-throughput investigations of the MCE. By applying this method to Gd, we obtain the MCE-induced temperature change of 1.84 ± 0.11 K under a modulation field of 1.0 T and modulation frequency of 0.5 Hz at a temperature of 300.5 ± 0.5 K, offering evidence that the LIT method gives quantitative results.

[a)] Present address: National Institute of Advanced Industrial Science and Technology, Nagoya 463-8560, Japan.

[b)] Present address: Jiangsu Key Laboratory of Advanced Micro & Nano Materials and Technology, School of Materials Science and Engineering, Nanjing University of Science and Technology, Nanjing 210094, China.

[c)] Corresponding author: UCHIDA.Kenichi@nims.go.jp

A solid-state refrigeration system based on so-called "caloric effects"[1] such as magneto-,[2-4] electro-,[5,6] baro-,[7] and elasto-[8] caloric effects has been proposed. Since these effects enable the construction of highly-efficient, safe, quiet, greenhouse-gas-free, and compact cooling systems, they are expected as next-generation refrigerators, especially for room-temperature applications. The caloric effects are driven by applying and releasing external fields, which are characterized by the isothermal entropy change and adiabatic temperature change $\Delta T_{ad}$. Especially, $\Delta T_{ad}$ is important because it is indispensable for efficient heat exchange from caloric materials to heat medium. To find and synthesize good caloric materials, systematic and high-throughput measurements of $\Delta T_{ad}$ are necessary.

However, conventional methods for measuring $\Delta T_{ad}$ have both merits and demerits. Typically, $\Delta T_{ad}$ is estimated by means of indirect methods. For example, $\Delta T_{ad}$ induced by the magnetocaloric effect (MCE) is characterized from the temperature dependence of the heat capacity at different magnetic fields,[9] which defines intrinsic $\Delta T_{ad}$ values. The $\Delta T_{ad}$ estimation, however, requires lengthy and time-consuming measurements under adiabatic conditions with varying magnetic fields and may include calculation errors due to the use of thermodynamic relations. Although $\Delta T_{ad}$ can also be measured directly by attaching temperature sensors to samples, the accuracy and versatility of such contact measurements are limited by unavoidable heat leakage between the sample and the sensors. Therefore, non-contact direct measurement methods for the MCE are required, and several non-contact methods were proposed. Otowski *et al.* first developed a photoacoustic-like detection method with lock-in systems in 1993 to evaluate $\Delta T_{ad}$ induced by small periodic magnetic fields, which enables systematic investigations of $\Delta T_{ad}$.[10-13] Döntgen measured the temperature dependence of the MCE in small volume and thin film samples by introducing a HgCdTe-infrared detector with lock-in systems.[14] While these lock-in-based methods provide high $\Delta T_{ad}$ sensitivity, they monitored the temperature only at a certain position of one sample, which limits throughput of the measurements. More recently, Ghorbani Zavareh *et al.* applied an infrared camera[15] to measure $\Delta T_{ad}$ induced by the pressure modulation. This method has a potential to obtain thermal responses from many samples at the same time, but the temperature resolution of such standard thermography measurements is much worse than the lock-in-based measurements.

To overcome the above problems, we apply the lock-in thermography (LIT) technique[16-20] to the measurements of the MCE. By using this technique, we realized systematic and quantitative evaluation of the thermal response of MCE materials induced by external magnetic fields, enabling easy, sensitive, and high-throughput measurements of the MCE. In the LIT-based MCE measurements, we apply a periodic magnetic field with the first harmonic amplitude $dH_{mod}$, modulation frequency $f$, and bias field $H_{bias}$ to MCE materials and extract thermal images oscillating with the field modulation using an infrared camera. Here, the temperature of the materials is modulated in response to $dH_{mod}$ due to the MCE as shown in Fig. 1(a). The obtained thermal images were transformed into the lock-in amplitude $dT_{mod}$ and phase $\varphi$ images by Fourier analyses; the $dT_{\mathrm{mod}}$ image shows the distribution of the magnitude of the field-induced temperature modulation and the $\varphi$ image shows the distribution of the time response of the output temperature modulation with respect to the input field modulation, which is determined by MCE properties as well as thermal diffusion. To systematically investigate the magnetic-field-dependent properties of MCE materials, we set the magnitude of $dH_{\mathrm{mod}}$ to small values, several tens of mT, and vary the bias field in the range of $0 < H_{\mathrm{bias}} < 1153$ mT. The total temperature change $\Delta T_{\mathrm{tot}}$ obtained under the magnetic field change between zero field and $H_{\mathrm{bias}}$ is estimated thorough the following equation:

$$\Delta T_{\mathrm{tot}} = \int_0^{H_{\mathrm{bias}}} \frac{dT_{\mathrm{mod}}}{dH_{\mathrm{mod}}} dH . \qquad (1)$$

In this study, we demonstrate the MCE measurements of Gd-Y alloys using the LIT technique. Note that the Gd-Y alloys are known to exhibit the MCE around room temperature with the Curie temperature $T_{\mathrm{C}}$ of around room temperature,[4] which makes the alloys suitable standard materials for the demonstration of the LIT-based MCE measurement method. A schematic of the experimental setup is shown in Fig. 1(b). The polycrystalline $Gd_{1-x}Y_x$ alloys with different $x$ values ($x$ = 0, 0.02, 0.04, 0.06, and 0.10) and pure polycrystalline Co were placed on a plastic plate base, where Co is a reference sample which shows no MCE around room temperature. The size of the samples is 2.5 mm × 2.5 mm × 0.4 mm. In order to minimize heat loss to surroundings, we dig a trench at the center of the plastic plate; the samples are attached to the plate only at the bottom edges [see Fig. 1(c)]. The $Gd_{1-x}Y_x$ alloys were prepared by arc melt and encapsulated in a quartz ampoule under helium atmosphere. Then, they were annealed at 1000 ºC for 20 h in order to homogenize them. Figure 2 shows the temperature dependence of the magnetization $M$ at 10 mT measured by SQUID VSM (Quantum Design Inc. MPMS3) for the $Gd_{1-x}Y_x$ alloys. The inset to Fig. 2 shows $T_{\mathrm{C}}$ for $Gd_{1-x}Y_x$ as a function of $x$, estimated by the Arrott plot.[21] We checked that $T_{\mathrm{C}}$ decreases with increasing Y concentration, consistent with reported values.[22] For the LIT measurements, the top surface (2.5 mm × 0.4 mm plane) of the samples was covered with black ink to ensure the infrared emissivity of > 0.95; see the steady-state infrared image for the samples shown in Fig. 1(d). The measurements were performed at the base temperature of 300.5 ± 0.5 K under an atmospheric pressure. The configuration of our LIT system (DCG Systems, ELITE) is detailed in Supplementary information of ref. 18.

In Fig. 3, we show examples of obtained $dT_{\mathrm{mod}}$ and $\varphi$ images under the several conditions of (a) $H_{\mathrm{bias}}$ = 275 mT, $dH_{\mathrm{mod}}$ = 24 mT, and $f$ = 0.5 Hz, (b) $H_{\mathrm{bias}}$ = 1127 mT, $dH_{\mathrm{mod}}$ = 10 mT, and $f$ = 5.0 Hz, and (c) $H_{\mathrm{bias}}$ = 51 mT, $dH_{\mathrm{mod}}$ = 2 mT, and $f$ = 25.0 Hz. The thermal responses from each sample can be immediately understood from these images. We confirmed that the $Gd_{1-x}Y_x$ alloys exhibit clear $dT_{\mathrm{mod}}$ signals and the magnitude of $dT_{\mathrm{mod}}$ decreases with increasing the Y concentration at the ambient temperature of 300.5 K, while there is no response from Co. The $\varphi$ images show that the temperature and field modulations for the Gd$_{1-x}$Y$_x$ alloys oscillate with almost the same phases; the temperature of the samples increases (decreases) when the magnitude of the external magnetic field increases (decreases). These results indicate that the temperature-modulation signals originate from the MCE and the $x$ dependence of $dT_{mod}$ for the Gd$_{1-x}$Y$_x$ alloys is consistent with the fact that the MCE is maximized in the vicinity of $T_C$. As is evident from Fig. 3(c), the LIT method realizes at least a few mK resolution for the MCE measurements, which cannot be achieved by the conventional steady-state thermography. Even sub-mK signals can be detected by increasing measurement time.

To demonstrate the quantitative capability of the LIT method for the MCE measurements, we focus on pure Gd, whose MCE properties are well known. Figure 4(a) shows $dT_{mod}/dH_{mod}$ as a function of $H_{bias}$ for Gd, obtained by averaging the LIT signals on the area defined by a white rectangle in Fig. 1 (d). The $dT_{mod}/dH_{mod}$ value shows a smooth rise with increasing $H_{bias}$, reaches maximum at around $H_{bias}$ = 0.4 T, and then gradually decreases with further increasing $H_{bias}$. This tendency reflects the fact that the $dT_{mod}/dH_{mod}$ is strongly related to the magnetic field dependence of $\partial M/\partial T$, consistent with the result in ref. 12. The $\varphi$ difference between $dH_{mod}$ and $dT_{mod}$ is less than 15 degrees, as shown in Fig. 4(b). The $\Delta T_{tot}$ values estimated from Eq. (1) as a function of $\Delta H_{tot}$, defined as a magnetic field change from zero to $H_{bias}$, at different lock-in frequencies are shown in Fig. 4(c). Here, the error bar was defined as $\pm 2\sigma$, where $\sigma$ is the standard deviation. The obtained value of $\Delta T_{tot}$ = 1.84 ± 0.11 K at $\Delta H_{tot}$ = 1.0 T and $f$ = 0.5 Hz for pure Gd was only 10 % lower than the reported value of $\Delta T_{ad}$ estimated by the conventional indirect method: the heat capacity measurement against the temperature at different fields [see Fig. 4(c)].[23] The difference can be attributed to small heat loss, from the sample to the plastic plate and air, and/or to the effective magnetic field of < 1.0 T decreased due to the demagnetization field determined by the sample shape. The small difference in $\Delta T_{tot}$ indicates that the LIT technique can be used for quantitative MCE measurements. As shown in the examples in Fig. 4, the LIT data provides systematic $H_{bias}$- and $f$-dependent properties of MCE materials.

Importantly, the LIT method allows us to measure the MCE of multiple materials at the same time, making this method attractive as a high-throughput measurement method of the MCE for new materials search. In Fig. 5(a), we show the $\Delta T_{tot}$ images under $\Delta H_{tot}$ = 1.0 T for all the samples at various values of $f$. From these images, we can easily find that $\Delta T_{tot}$ decreases with increasing $x$ of Gd$_{1-x}$Y$_x$. We can extract a variety of the data sets from the $\Delta T_{tot}$ images for all the samples; in Fig. 5, we show the $\Delta T_{tot}$ values as functions of $f$ at $\Delta H_{tot}$ = 1.0 T [Fig. 5(b)], $\Delta H_{tot}$ at $f$ = 0.5 Hz [Fig. 5(c)], and $x$ at $\Delta H_{tot}$ = 1.0 T [Fig. 5(d)]. Interestingly, when $f$ > 10 Hz, the magnitude of $\Delta T_{tot}$ gradually decreases with increasing $f$ and this behavior becomes prominent with increasing $x$, implying that the suitable operation frequency changes depending on $x$ in the Gd$_{1-x}$Y$_x$ alloys probably because of the difference in thermal properties.

As demonstrated here, the simultaneous measurements of many materials are useful for clarifying and comparing MCE properties of the materials.

Finally, we discuss the usefulness of the LIT-based MCE measurement method. We believe that this method has two major merits. One is to obtain systematic data set in wide operation-frequency and magnetic-field ranges, which is helpful for optimizing refrigeration cycles to maximize the cooling power based on the entropy-temperature diagram drawn by the data set at different operation frequencies. Moreover, the data set for many samples can be obtained at the same time by using the LIT-based method, which accelerates a material search for better MCE materials. Therefore, the high-throughput measurement method developed here will strongly contribute not only to evaluating MCE properties for samples with various compositions but also to screening new series of materials. The other is to obtain dynamic thermal responses including thermal diffusion and exchange to heat medium, which can be visualized as $\varphi$ images. Especially, the LIT data set is valuable for the materials with first-order magnetic phase transitions and larger MCE at around room temperature, such as $Gd_5Ge_2Si_2$,[24] $Fe(La,Si)_{13}$,[25,26] FeRh,[27,28] $(MnFe)_2(SiP)$,[29,30] and Ni-Mn-In-Co,[31] since thermal and magnetic hysteresis cannot be neglected[31,32] in thermodynamic cycles. Note that the data set of thermal response induced by the external field change with high frequency (>10 Hz) modulation is significant from the viewpoint of practical MCE applications. This is because the operation frequency of > 10 Hz is currently required in order to enhance the cooling power, although the operation frequency for existing magnetic refrigeration systems is around 0.1-4.0 Hz.[33] The LIT method can cover the frequency range much higher than that required for MCE applications.

In summary, we have demonstrated that the lock-in thermography (LIT) technique enables systematic, quantitative, and high-throughput investigations of the magnetocaloric effect (MCE). The LIT realizes non-contact direct measurements of the adiabatic temperature change induced by the MCE, free from any possible calculation errors arising from indirect measurements based on thermodynamic relations. The LIT can evaluate the maximum cooling power for the magnetic refrigeration by drawing a thermodynamic cycle, although it remains to be demonstrated. More significantly, the measurement technique developed here is potentially applicable not only to the MCE but also to other caloric effects, such as electro-, baro-, and elasto-caloric effects, by replacing periodic external magnetic fields with electric fields, pressure, and strain. Therefore, we anticipate that the LIT method will accelerate materials research towards high-efficient solid-state refrigeration based on the caloric effects.

**Acknowledgement**
The authors thank H. Sepehri-Amin for valuable discussions. This work was supported by Grant-in-Aid for Scientific Research (A) (JP15H02012) from JSPS KAKENHI, Japan. One of the authors, X. F. Miao, acknowledges LGE for financial support.

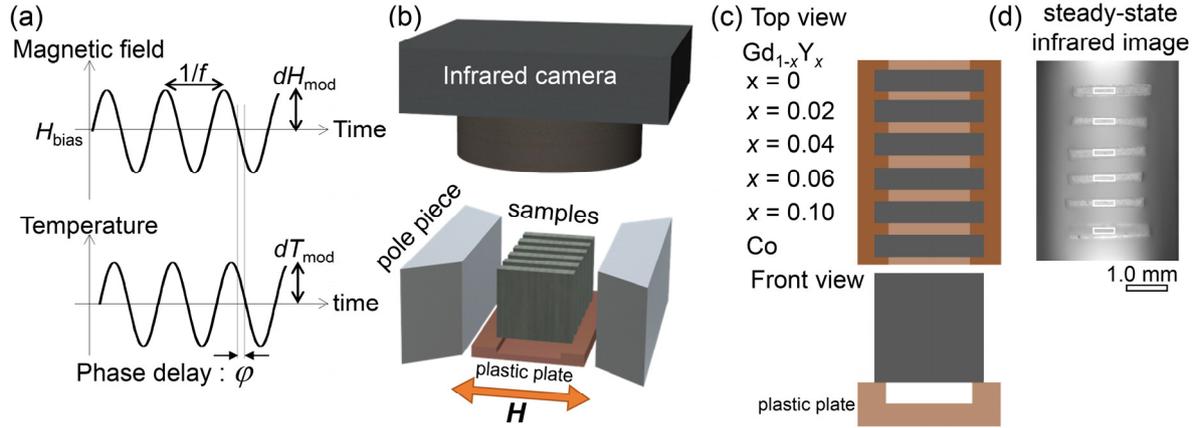

FIG. 1. MCE measurements based on LIT. (a) Input and output of the LIT-based MCE measurements. The input is a periodic magnetic field with the frequency $f$, amplitude $dH_{\text{mod}}$, and bias $H_{\text{bias}}$. The output is a periodic temperature modulation induced by the MCE with the frequency $f$, amplitude $dT_{\text{mod}}$, and phase delay $\varphi$. (b) A schematic illustration of the experimental setup. 6 samples, $Gd_{1-x}Y_x$ alloys ($x$ = 0, 0.02, 0.04, 0.06, and 0.10) and Co, were put on a plastic plate between the pole pieces of an electromagnet. The thermal response from the top surface of the samples is detected by the infrared camera. (c) Top and front views of the sample setting. The 6 samples were bridged on the plastic plate with a trench by using superglue in order to suppress the heat leakage to the base. (d) Steady-state infrared image of the samples. The MCE signals for the samples were evaluated by averaging the data in the areas defined by white rectangles.

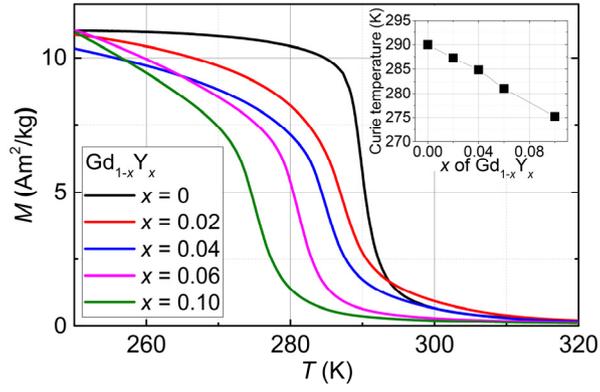

FIG. 2. Temperature dependence of the magnetization $M$ for the $Gd_{1-x}Y_x$ alloys ($x$ = 0, 0.02, 0.04, 0.06, and 0.10) at 10 mT with the temperature rising rate of 2 K/min. The Curie temperature $T_C$ was determined by the Arrott plot.

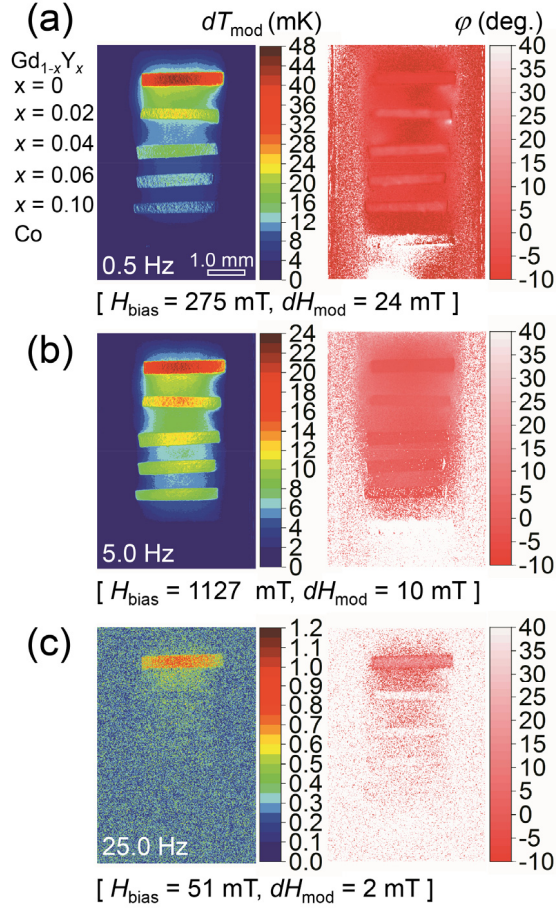

FIG. 3. $dT_{mod}$ and $\varphi$ images for the Gd$_{1-x}$Y$_x$ and Co samples under the conditions of (a) $H_{bias}$ = 275 mT, $dH_{mod}$ = 24 mT, and $f$ = 0.5 Hz, (b) $H_{bias}$ = 1127 mT, $dH_{mod}$ = 10 mT, and $f$ = 5.0 Hz, and (c) $H_{bias}$ = 51 mT, $dH_{mod}$ = 2 mT, and $f$ = 25.0 Hz.

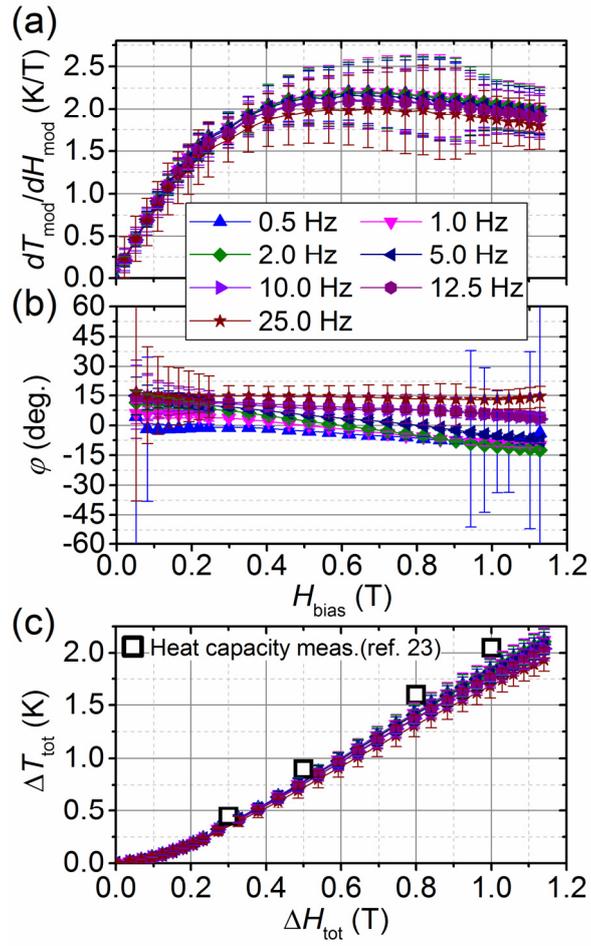

FIG. 4. Systematic data set for Gd. (a), (b) $H_{bias}$ dependence of $dT_{mod}/dH_{mod}$ and $\varphi$ for various $f$ values. (c) $\Delta H_{tot}$ dependence of $\Delta T_{tot}$ for various $f$ values. The black rectangles correspond to the adiabatic temperature change estimated from the conventional indirect measurements shown in ref. 23.

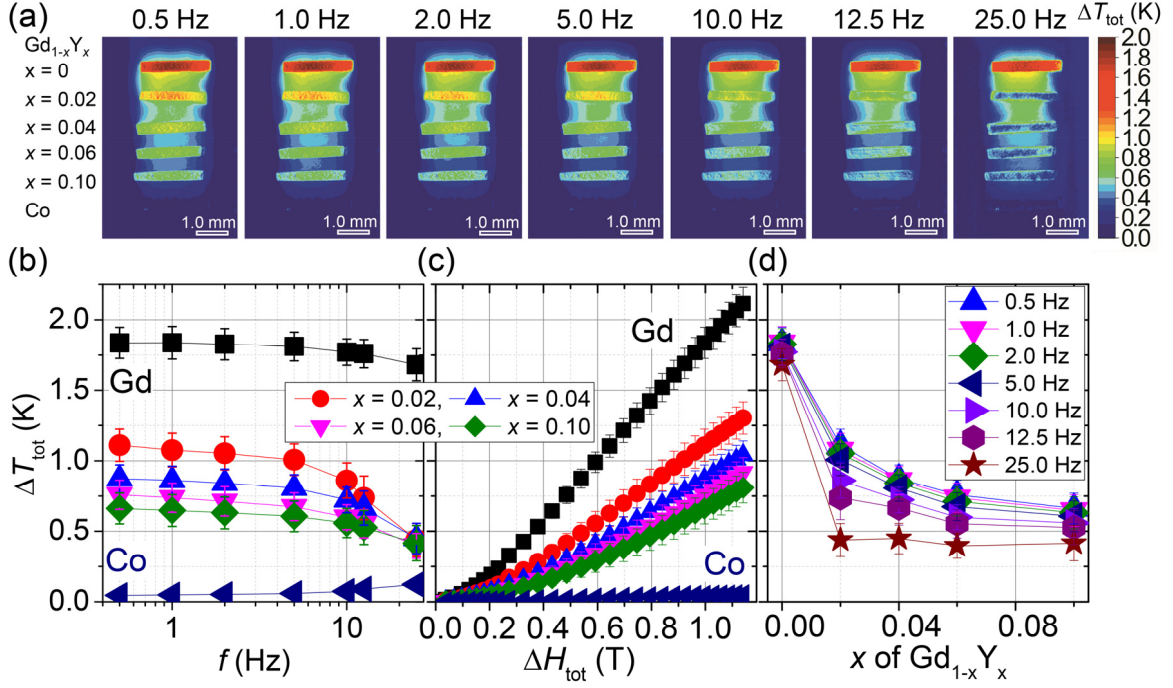

FIG. 5. Dependence of MCE properties on material compositions. (a) $\Delta T_{tot}$ images for the $Gd_{1-x}Y_x$ and Co samples at $\Delta H_{tot} = 1.0$ T for various $f$ values. (b) $f$ dependence of $\Delta T_{tot}$ for the $Gd_{1-x}Y_x$ and Co samples at $\Delta H_{tot} = 1.0$ T. (c) $\Delta H_{tot}$ dependence of $\Delta T_{tot}$ for the $Gd_{1-x}Y_x$ and Co samples at $f = 0.5$ Hz. (d) $x$ dependence of $\Delta T_{tot}$ for the $Gd_{1-x}Y_x$ samples at $\Delta H_{tot} = 1.0$ T for various $f$ values.